\documentclass[aps,pre,twocolumn]{revtex4-1}  

\usepackage{amsmath}    
\usepackage{amssymb}
\usepackage{graphicx}   
\usepackage{color}
\usepackage{subfigure}
\usepackage{cases}
\usepackage{epstopdf}
\newcommand{\beq}{\begin{equation}}
\newcommand{\eeq}{\end{equation}}

\begin{document}

\title{Critical exponents in zero dimensions}

\author{A. Alexakis,   F. P\'etr\'elis}
\affiliation{Laboratoire de Physique Statistique, Ecole Normale Sup\'erieure, CNRS, 24 rue Lhomond, 75005 Paris}


\date{\today}

\begin{abstract}
In the vicinity of the onset of an instability, we investigate the effect 
of colored multiplicative noise on the scaling  of the  moments of the 
unstable mode amplitude. 
We introduce a family of zero dimensional models for which we can 
calculate  the exact value of the critical exponents $\beta_m$
for all the moments. The results are obtained through  asymptotic expansions that use the distance to onset as a small parameter. 
The examined family displays a variety of behaviors of the critical exponents that includes anomalous
exponents: exponents that differ from the deterministic (mean-field) prediction, and multiscaling: non-linear dependence of the exponents on the order of the moment. 
%
 
 



\end{abstract}

\maketitle

\section{Introduction}

Critical exponents are usually introduced in the context of continuous phase transitions at equilibrium.  
The order parameter (for instance the magnetization for a system of spins or the density difference between the phases 
for the liquid-gas critical point)   depends on the distance from the critical point as a power-law. The exponent of the 
power-law, traditionally named $\beta$, is one of the critical exponents of the system. Mean-field approximations simplify 
the analytical approach  and allow to calculate  $\beta$. In this case simple 
rational values are obtained that depend on the nonlinearity (itself usually constrained by the symmetries) of the system. 
 However, because of thermal fluctuations the mean-field results are not always correct for low spatial dimensions. 
In particular $\beta$ can take non-mean-field values \cite{Kada}. 
Apart from the few cases in which exact analytical results exist, renormalization
methods are used from which $\beta$ is  obtained as a series in the critical dimension  (the dimension above which mean-field
results apply) minus the spatial dimension \cite{wilson}.  The importance of spatial dimensionality appears also clearly in 
the fact that equilibrium phase transitions do not occur when the spatial dimension is too small and short range interactions 
are considered. For continuous phase transitions at equilibrium, thermal  fluctuations, nonlinearities and spatial 
variations must be taken into account.

In these systems at equilibrium, fluctuations are additive terms in the equation for the order parameter.  
 In out of equilibrium systems, fluctuations can be coupled differently to the order parameter
and in several models, multiplicative noise is considered. 
Then, the analogous of phase transition can occur 
even when space is not taken into account and only the time
evolution of a finite number of modes is considered. 
Since spatial dimensions are not taken into account we refer to these models as zero dimensional. 
The simplicity of zero dimensional models allows a much more thorough analytical investigation and in a few cases
the calculation of the full probability distribution function (pdf) is possible. This is the direction that we pursue in this work.

We introduce a family of zero dimensional models for which we can calculate (in the small deviation from criticality limit)
the stationary pdf  of the system and thus we can obtain the exact value of the critical exponents $\beta_m$ for all the moments. 
Despite the simplicity of the models the results show a rich behavior.  The critical exponents can differ from their deterministic values and in some cases vary continuously with the system parameters.    In the later case a non-linear dependence of the exponents on the order of the moment
is observed and thus the system displays multiscaling.

In section \ref{Sec_model}, we present what is known in the deterministic limit and in the case of a white noise.  
The family of models under study is presented in section \ref{Sec_modelb}. 
In section \ref{Sec_nrm} and section \ref{Sec_Anm}, we present two particular cases the second of which 
results in anomalous exponents. 
In section \ref{Sec_Hrs}, our results are interpreted based on a heuristic arguments 
from which the value of the critical exponents is understood. 
We conclude in the last section. In \cite{prl}, we reported on the value of the exponent of the first moment obtained for certain values of the parameters. 
The associated asymptotic expansion is presented in detail here together with several new expansions that are valid for other 
parameter values. Overall we are now able  to calculate the whole set of exponents of all orders of the described model.

\section{Zero dimensional bifurcations}\label{Sec_model}

We consider  the evolution of an order parameter, $x$, which is a function only of  time $t$. It satisfies 
the Langevin equation 
\begin{equation}
\dot{x}  = \mu\, x - |x|^{n}\,x +  \xi\,x,
\label{eq1}
\end{equation}
where $\mu$ is the parameter that controls the instability, $n>0$ characterizes the nonlinearity 
(for instance $n=2$ for cubic nonlinear terms, $n=4$ for quintic ones) and $\xi$ represents random fluctuations with zero mean (noise). 
 From now on, in the case that $\xi$ is white noise, 
we use the Stratanovich interpretation \cite{vankampen}. We note that  the solution $x(t)$ conserves its 
sign and we  thus restrict only to positive values 
for $x$. We are going to characterize the behavior of $x$ using its moments evaluated in the long time limit 
that we write  here as $\langle x^m \rangle$,
where $\langle \cdot \rangle$ stands for average over the realizations of the noise. 
The stability of the $x=0$ solution is determined  by calculating  the value of the growth rate 
$\gamma = \langle \dot{x}/x \rangle$ for the linear system \cite{arnold}. 
Here $\gamma=\mu$ 
thus the onset of the instability is given by $\mu=0$. For $\mu<0$ the 
only attracting solution of the system is $x=0$ and thus all moments are zero.
For positive values of $\mu$, $\langle x^m \rangle$ takes non-zero values whose amplitude has a power-law 
dependence on $\mu$: $\langle x^m \rangle  \propto \mu^{\beta_m}$. 
The exponents of these power-laws $\beta_m$  are of primary interest in this work.  
Explicitly we define 
\begin{equation}
\beta_m \equiv \lim_{\mu\to 0}  \log (\langle x^m \rangle) / \log(\mu).  
\end{equation}

In the deterministic limit ($\xi=0$) and 
for positive $\mu$ the long time solution satisfies
$\lim_{t\to\infty}x^m=\mu^{m/n}$. The critical exponents are thus  $\beta_m=\frac{m}{n}$. 
We will refer to this scaling as mean-field  or deterministic scaling. 

\begin{figure*}[t!]
\begin{center}
\centerline{
\includegraphics[width=6.0cm,height=2.7cm]{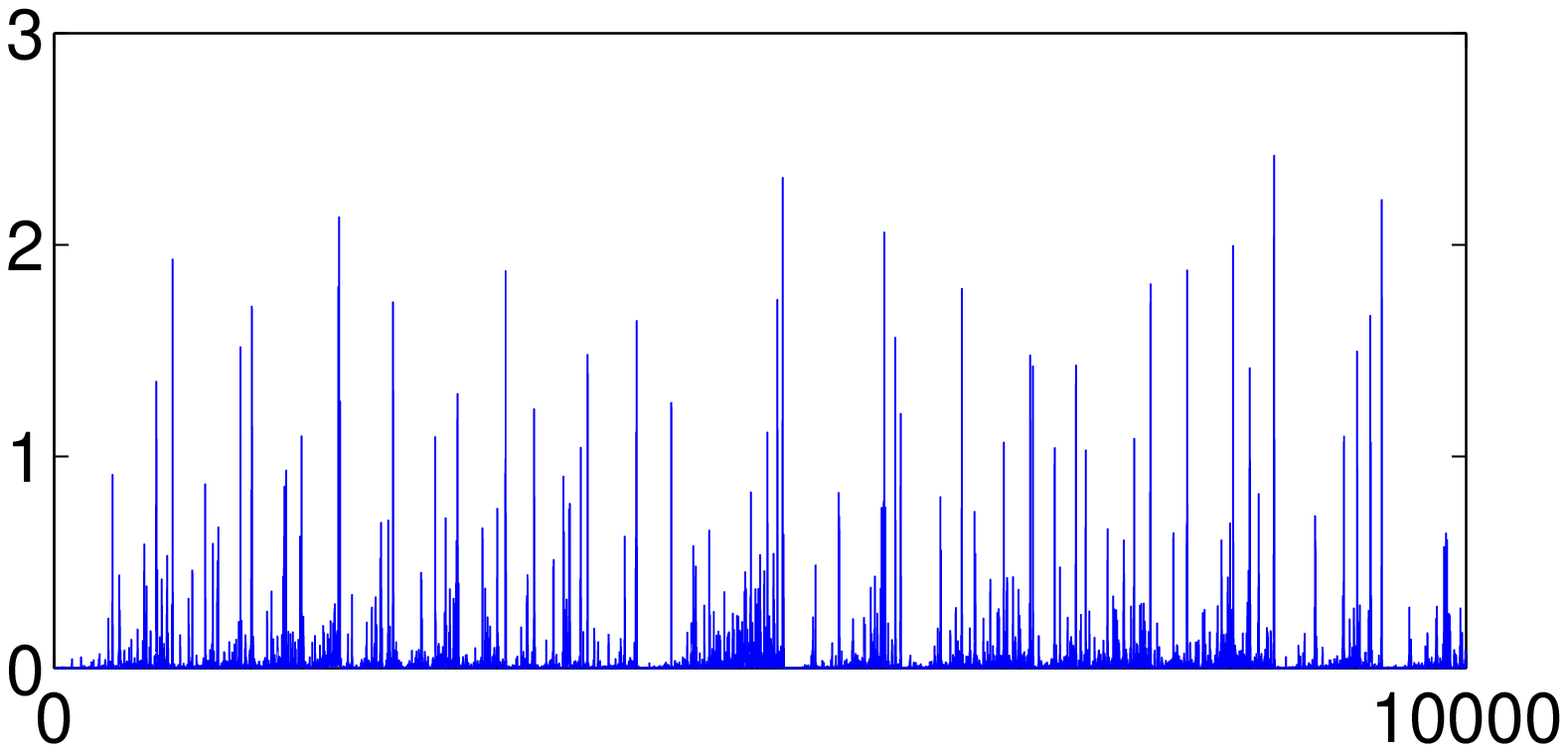}    
\includegraphics[width=6.0cm,height=2.7cm]{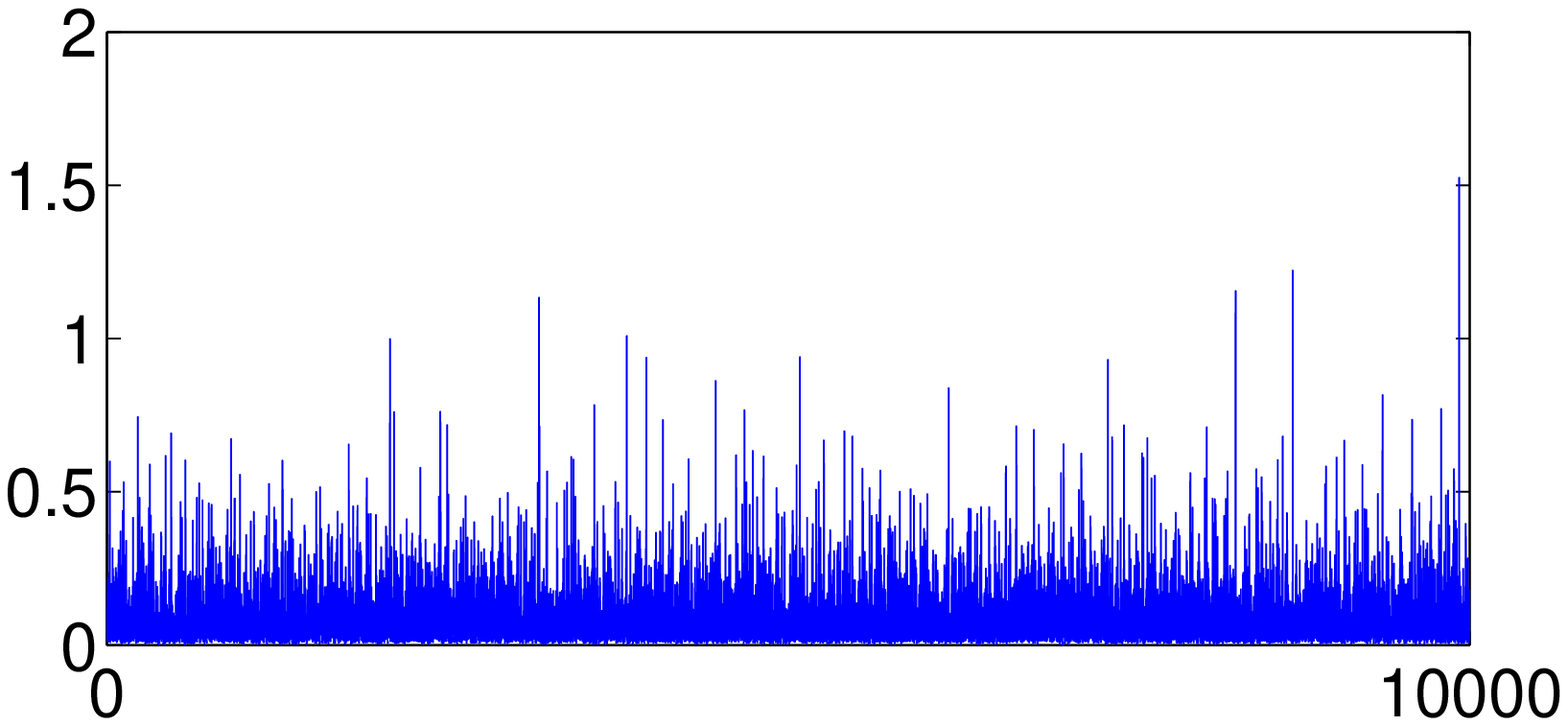}    
\includegraphics[width=6.0cm,height=2.7cm]{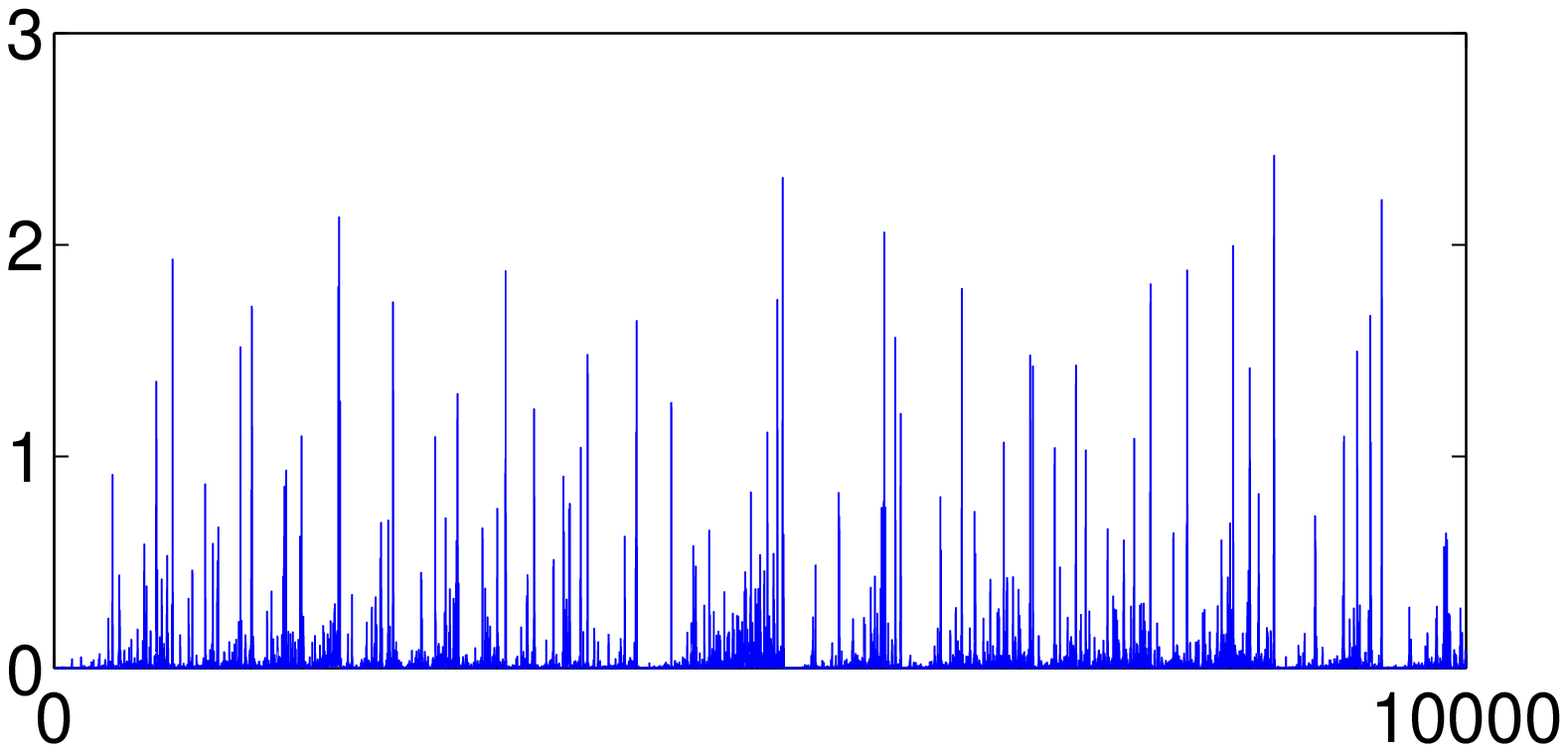}         }
\centerline{  (a) \hspace{5.5cm} (b) \hspace{5.5cm} (c)                }
\vspace{-0.5cm}
\end{center}
\caption{\label{figtime} Time series of the solution of eq. (\ref{eqdepart}) for $\mu=0 .01$. 
Top:  linear scale and bottom in log scale. 
a) $F=0$ corresponding to a white noise;  
b) $F=\gamma Y^2/2$, Ornstein-Uhlenbeck noise with $\gamma=1.5$; 
c)  $F=\nu |Y|$, with $\nu=0.75$. Note the differences in the $y$-coordinate values.}
\end{figure*}

A second well-studied limit is obtained when $\xi$ is a Gaussian  delta-correlated noise,  {\it i.e.} 
$\langle \xi(t)\xi(t') \rangle= 2 \delta(t-t')$. Then the stationary probability distribution function (pdf) of $x$, 
$P(x)$, satisfies the one dimensional Fokker-Planck Equation 
\beq
\partial_x (\mu x - x^{n+1})P + \partial_x x \partial_x x P  =0.
\eeq
Its solution is given by 
\beq
P(x)= \frac{n^{1-\mu/n} }{\Gamma\left(\mu/n\right)} \, x^{\mu-1}  e^{- x^n/n}
\eeq 
where the normalization condition $\int P(x)\,dx =1 $ has been used.
The moments can then be calculated as
\beq
\langle x^m \rangle = \int P \,x^m dx\,
\eeq
that, due to the singularity of $P$ at $x=0$, result to $\langle x^m \rangle  \propto  \mu$ in the small $\mu$-limit. 
We thus have $\beta_m=1$ for all moments $m$. 
%
This property is an effect of the noise on the dynamics of $x$ close to the onset. 
%
Indeed, the time series alternates between phases where the value of $x$ 
is either large and nonlinearities are important  (on-phases) or it fluctuates close to zero (off-phases) 
\cite{on-off, on-off1, on-off2}. 
This behavior is called on-off intermittency and an example of time series is displayed in fig. \ref{figtime}. 
%
%
The mean duration of the off-phases, say $T_{_{OFF}}$ can be estimated by considering the evolution of 
$z=\log(x)$ for which eq. (\ref{eq1}) is written as 
\beq
\label{LG2}
\dot{z}  = \mu\, +  \xi\, - e^{nz} \,.
\eeq
Thus $z$ displays Brownian motion with a small drift ($\mu$) when $x\ll1$ while it is repelled towards 
the origin $x=0$ by the nonlinearity when $x$ is order 1. As a result the duration of the off-phases diverges as 
$T_{_{OFF}}\sim \mu^{-1}$ 
while the duration of the on-phases 
remains finite. During the on-phases, $x$ achieves finite values, say $x_{_{NL}}$ that do not depend on $\mu$ (in the small 
$\mu$ limit). An estimate of the moments is given by 
$\langle x^m \rangle\simeq T_{_{ON}} {x^m_{_{NL}}}/ (T_{_{ON}}+T_{_{OFF}})$ 
from which the linear dependence of the moments on $\mu$ is recovered \cite{seb1}.

\section{Bifurcations in the presence of colored Noise}\label{Sec_modelb}


For  colored noise, it has been shown that the regime of on-off intermittency is controlled by the value of the noise 
spectrum at zero frequency, $D=\int_0^{\infty} \langle  \xi(t) \xi(0) \rangle dt$ \cite{wien}. As long as $D$ is non zero, the behavior 
for very small $\mu$ is on-off intermittency \cite{seb1}. In what follows, we examine the properties of the bifurcation when the 
  noise has vanishing  spectrum at zero 
frequency.

We consider the family  of models 
\begin{eqnarray}
\dot{x} & =& \mu x - x^{n+1} + x [\xi -  F_y(y)]\,,\nonumber \\
\dot{y} &= &[\xi - F_y(y)]
\label{eqdepart}
\end{eqnarray}
where $\xi$ is a Gaussian white noise, $\langle \xi(t)\xi(t') \rangle= 2 \delta(t-t')$. We have introduced the potential $F(y)$ which is a function of $y$ only,
and $F_y$ its first derivative. 

The amplitude of the order parameter, $x$, undergoes a bifurcation at $\mu=0$ and is subject to a multiplicative noise $\dot{y} = \xi - F_y(y)$.  
Provided the stationary distribution of  $y$ has a  finite second moment, the spectrum of $\dot{y}$ vanishes at zero frequency. Indeed, for initial conditions $y(0)=0$, we can write
\begin{equation}
\frac{d y^2}{dt}=2 y \dot{y}=2 \dot{y}(t) \int_0^t \dot{y}(t') dt'\,.
\end{equation}
By averaging over the realizations and taking the long time limit, the last expression is the integral of the autocorrelation of $\dot{y}$. This is also its spectrum at zero frequency using the Wiener-Kintchin theorem. At long time, if the second moment of $y$ tends to a constant, then $\dot{y}$ has vanishing spectrum at zero frequency. 

The stationary pdf for $y$,  $\Pi(y)$, satisfies the equation 
$\mathcal{L}_0 \Pi=0$, where the linear operator $\mathcal{L}_0$ is defined as
\beq
\label{lin_op}
\mathcal{L}_0 \Pi \equiv 
\partial_y [ (\partial_y F(y)) \Pi ] + \partial_y^2 \Pi\,, 
\eeq
and the normalization condition $\int \Pi dy=1$ is assumed.
Equation $\mathcal{L}_0 \Pi=0$ has the solution
\beq
\Pi(y) = Exp[ -F(y) ]/N\,,
\eeq
where $N$ is a normalization constant.

To isolate the noise term we make the following transformation $w=\log(x)-y$. 
The Langevin equation becomes
\begin{eqnarray}
\dot{w} & =& \mu  - e^{n(w+y)}\,, \\
\dot{y} &= &\xi - F_y.
\label{wyeq}
\end{eqnarray}

The Fokker-Planck equation for the stationary joint pdf $P(x,y)$ in these coordinates then reads 
\beq
\partial_w (\mu -  e^{n(w+y)} ) P = \partial_y [ F_y P ] + \partial_y^2 P\,=\mathcal{L}_0 P.
\label{EFP}
\eeq
Solving the partial differential equation (\ref{EFP}) for all values of the parameters is out of reach. 
Since we are interested in the critical behavior, we successively introduce various asymptotic approaches in 
order to determine the critical exponents.

\section{ Steep potential - Recovery of the Deterministic scaling}
\label{Sec_nrm}
\subsection{The general case}

The top panel of 
figure \ref{fig2} shows the location in phase space of $10^3$ trajectories for different values
of $\mu$ (different colors) obtained by the numerical integration of the Langevin equations \ref{eqdepart} for a steep potential that is  defined in section IV.B.
It can be seen that as $\mu$ becomes smaller the distribution is concentrated around a value of $w=w_0$
that depends on $\mu$.  
Inspired from the numerical results and
without any assumptions yet on the functional form of $F$, we make the following change of variables
$w=w_0 + \ln(\mu) /n + u \sqrt{\mu } $ where $w_0=\ln(X_0)$ is a constant that will be set by the expansion. 
The Fokker-Planck equation then reads
\beq
\mathcal{L}_0 P =
\sqrt{\mu} \partial_u ( 1 - X_0^n  e^{n( \sqrt{\mu} u+y)} ) P. 
\eeq
We expand the p.d.f. as
$P=P_0+\sqrt{\mu} P_1 + \mu P_2 +\dots $.

To lowest order we obtain the equation for the stationary distribution in $y$
\beq
{\cal L}_0 P_0 =0\,.
\eeq
The solution of which can be written as
\beq
P_0 = 
A(u) \Pi_0(y)= 
A(u) e^{-F}\,,
\eeq
where the amplitude $A(u)$ is left undetermined. 
To next order we have
\beq
{\cal L}_0 P_1 =\partial_u ( 1 -  X_0^n e^{ny} ) P_0\,. 
\eeq

Integrating this equation over $y$ provides us with a solvability condition. More precisely,
since ${\cal L}_0 P_1$ is a total gradient, 
integration over $y$ makes the left hand side equal to zero and we are thus left with 
\beq
0 = \int_{-\infty}^{+\infty} ( 1 -  X_0^n e^{ny} ) e^{-F} dy,
\eeq
This condition determines the value of $X_0$ to be
\beq
X_0 = \left( \frac{\int_{-\infty}^{+\infty} e^{-F} dy}{\int_{-\infty}^{+\infty}  e^{ny-F}  dy} \right)^{1/n}.
\label{eqflux}
\eeq
We can then solve for $P_1$ and we obtain
\begin{eqnarray} 
P_1 &=& A_u e^{-F} 
            \int_0^y            e^{F''} 
            \int_{\infty}^{y''} (1-  X_0^n e^{ny'})e^{-F'} dy' dy''  \,,\nonumber \\
&=& A_u \Pi_1(y) \,,
\end{eqnarray}	
where $F'$ and $F''$ denotes that the function $F$ depends on the variables 
$y'$ and $y''$ respectively. 
\noindent
To second order we have
\beq
{\cal L}_0 P_2=\partial_u \left[ ( 1 -  X_0^n e^{ny} ) P_1 -  u X_0^n e^{ny} P_0 \right]  \,.
\eeq
Using again the solvability condition, we integrate over $y$ and  obtain
\beq
   A_u \int_{-\infty}^{+\infty} (1-X_0^ne^{ny}) \Pi_1 dy  -  A u X_0^n \int_{-\infty}^{+\infty} e^{ny-F } dy =0\,,
\eeq
that leads to
$  A= e^{-\delta   u^2 }  $
with
\beq
\delta = \frac{1}{2} X_0^n \int_{-\infty}^{+\infty}  e^{ny-F(y) } dy \Big{/}  \int_{-\infty}^{+\infty}  (X_0^ne^{ny}-1) \Pi_1(y) dy\,.
\label{eqlambda}
\eeq
Denoting  $Q(y)=\int_{-\infty}^y (X_0^ne^{ny'}-1) e^{-F'} dy'$, we can show by integration by parts 
that
\[ \int _{-\infty}^{+\infty} (X_0^ne^{ny}-1) \Pi_1(y) = \int_{-\infty}^{+\infty}  Q^2e^F dy >0 \,,\]
thus $\delta>0$.
The zeroth order solution then becomes
\beq
P_0 = \frac{1}{N} e^{-\delta u^2 -F(y)} \,,
\eeq
where $N=\sqrt{\frac{\pi}{\delta}} \int_{-\infty}^{+\infty}  e^{-F} dy$. 
The moments can  be calculated as
\beq
\langle x^m\rangle = \frac{\mu^{m/n} X_0^{m/n}}{N} \int_{-\infty}^{+\infty}  e^{my+m \sqrt{\mu} u -\delta u^2 -F(y) } dy d u \,.
\eeq
The integral and $N$ are independent to first order in $\mu$. Provided the integral and $X_0$ are finite,  the moments follow the scaling
$\langle x^m\rangle \propto \mu^{m/n}$. Therefore, we recover the deterministic exponents $\beta_m=m/n$.

\subsection{An example: $y$ is the  Ornstein-Uhlenbeck process}

\begin{figure}[h!]
\begin{center}
\includegraphics[width=8cm,angle=0]{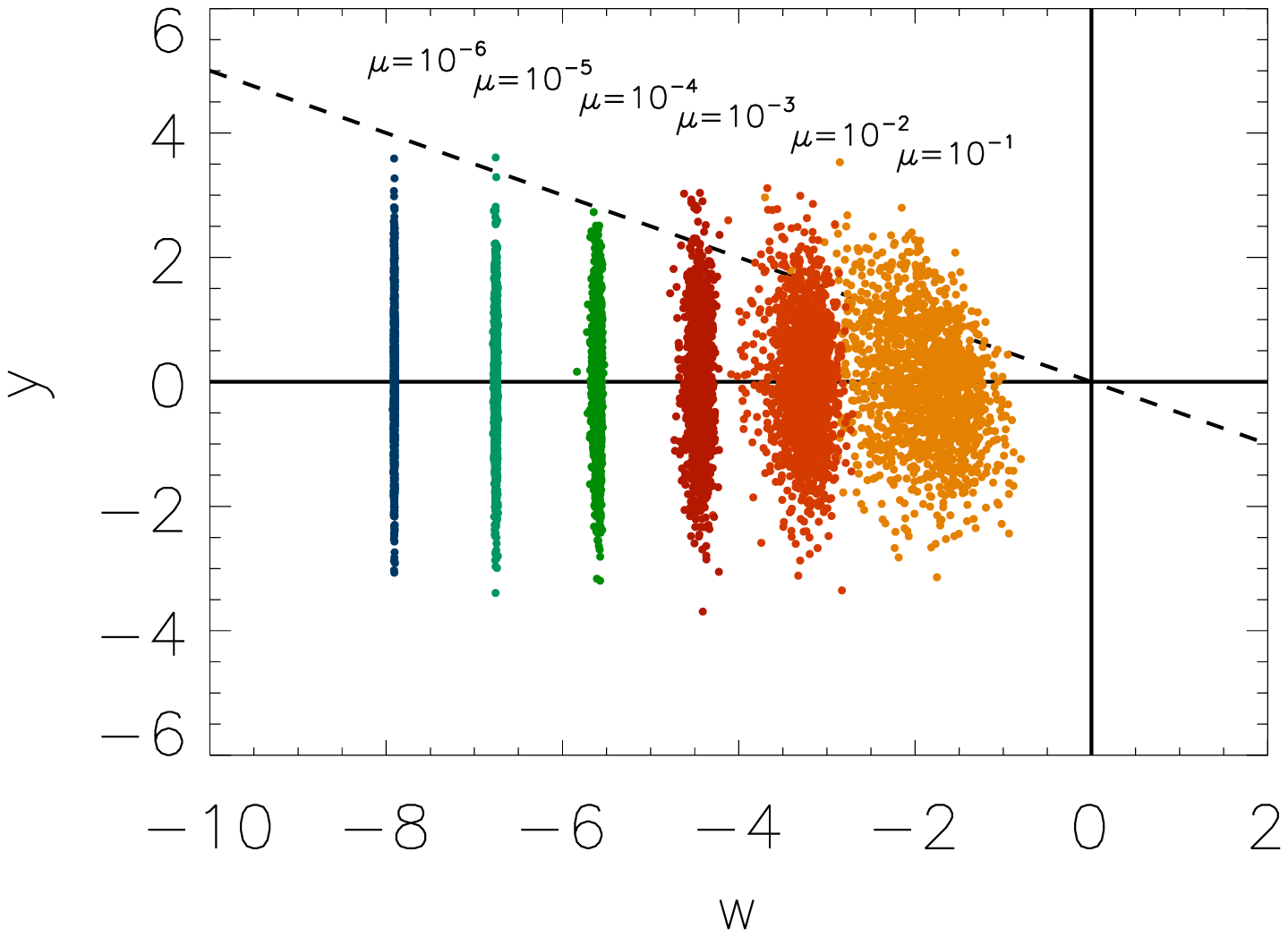}
\includegraphics[width=8cm,angle=0]{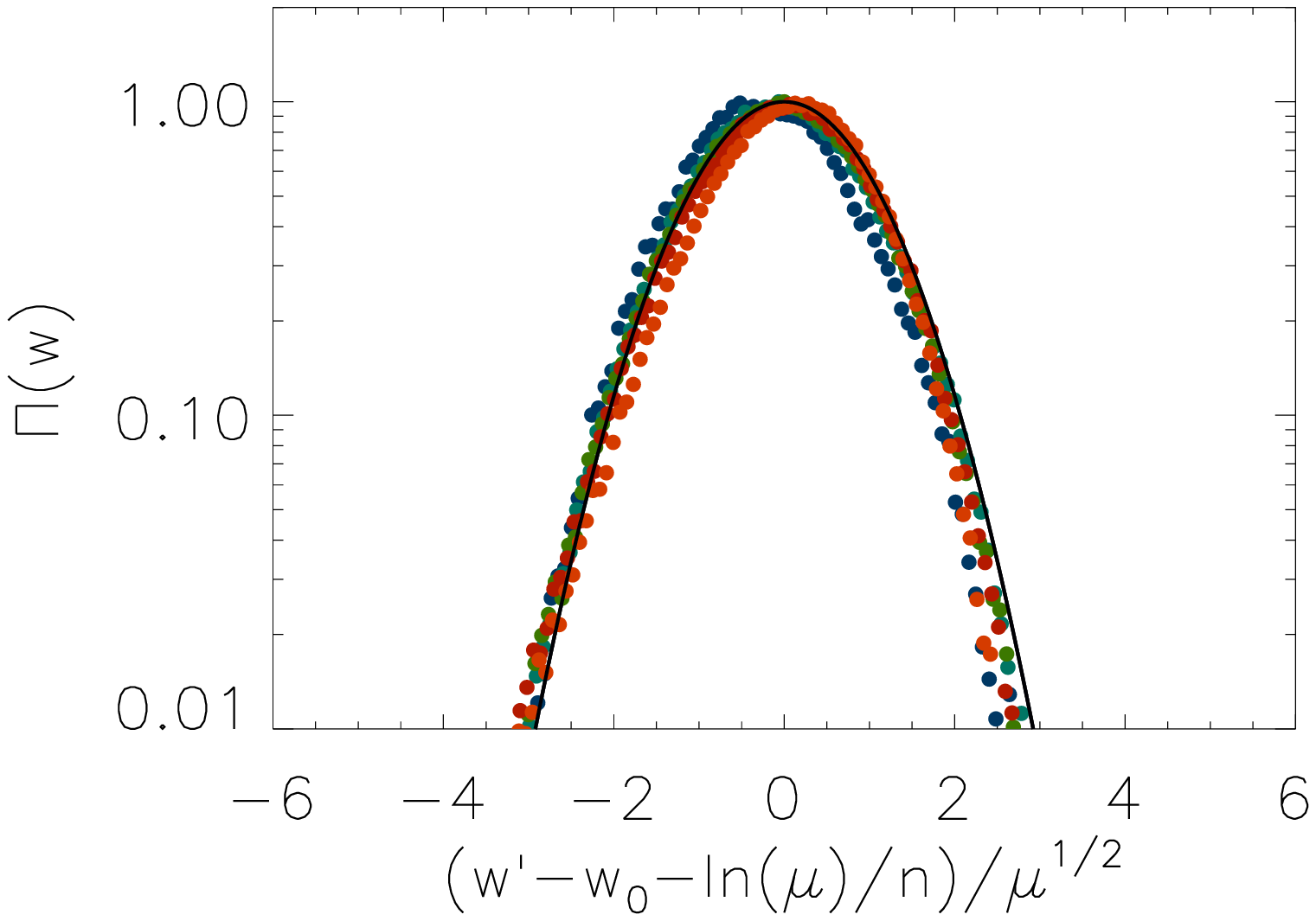}
\caption{\label{fig2} 
                      Top panel: Position in phase-space of  $10^3$ realizations of the numerical solution 
                      of the Langevin Equation \ref{eqdepart} with $F=\frac{1}{2}\gamma y^2$, and $\gamma=2$.  
                      Different shades/colors (online) correspond to different values of $\mu$ as indicated. The dashed line corresponds to $w+y=0$ where the nonlinear term is of order one. 
                      Lower panel: The marginal probability $\Pi(w)=\int P(y,w)dy$ from the numerical investigation
                      (dots) and the analytical prediction (solid line). Shades and colors are the same as in the top panel.    }
\end{center}
\end{figure}

A simple potential for which the former expansion is valid consists in  $F(y)=\frac{1}{2}\gamma y^2$. 
This case corresponds to the  Ornstein-Uhlenbeck process for the variable $y$ \cite{vankampen}.
Time series of $x$ are presented in fig. \ref{figtime}. As mentioned, we show in fig. \ref{fig2} the position of several trajectories in phase-space from which the concentration of the p.d.f around the value $w_0$ appears clearly.   In this case the p.d.f. at lowest order becomes
\begin{equation}
P_0 = \frac{1}{\sqrt{\pi}} \exp\left[ -\frac{\delta}{\mu} (w-w_0-\frac{1}{n}\ln(\mu))^2 - \frac{1}{2}\gamma y^2 \right]\,,
\label{eqP0}
\end{equation}
with $w_0 =\log X_0= -n/(2 \gamma)$ and $\delta$ is given by  the integral (\ref{eqlambda}). 
Note that $X_0$ takes very small values when $\gamma$ is small which requires high accuracy when solving numerically the 
Langevin equation (\ref{eqdepart}). In fig. \ref{fig2}, we present the marginal probability $\int P(\omega,y) dy$ as a 
function of $(w-w_0)/\mu^{1/2}$ for different values of $\mu$ and  $\gamma=1$. 
The numerically computed pdf agree well with the theoretical expression \ref{eqP0}.

\subsection{Crossover: Two non-commutative limits}

\begin{figure}[h!]
\begin{center}
\includegraphics[width=8cm]{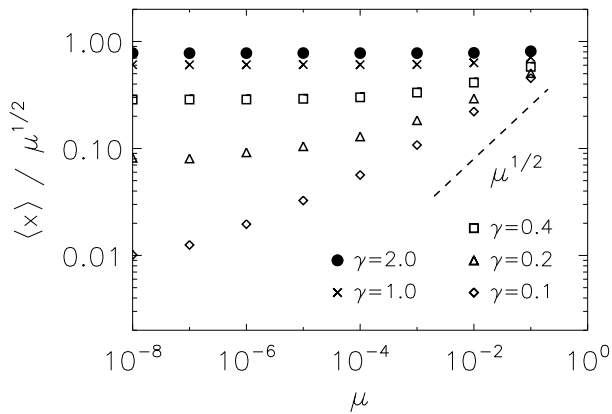}
\caption{First moment as a function of $\mu$ obtained from the numerically computed solutions of eq. \ref{eqdepart} for $F=\gamma y^2/2$ and five different values of $\gamma$. The first moment has been rescaled by the deterministic scaling $\mu^{1/2}$. \label{fig3}  }
\end{center}
\end{figure}%

Setting $\gamma=0$ in eq. (\ref{eqdepart}), we recover eq. (\ref{eq1}) with a white noise and this is the case in which on-off intermittency takes place. 
In this case, the exponents are $\beta_m=1$. This is at odd with the deterministic scaling $\beta_m=m/n$ that we have 
predicted for non-zero $\gamma$ (but possibly arbitrarily small).  

Some insight on this problem of exchange of limits can be obtained by 
investigating 
the small $\gamma$ limit (or small amplitude of $F$ in general).
In this case, we write  $P=P_0+ \gamma^{1/2} P_1 + \gamma P_2 +\dots$
and use the fast $z=w+y=\ln(x)$ and slow $Y=\gamma^{1/2} y$ variables. 
The Fokker-Planck in these coordinates reads
\[
\partial_z (\mu -  e^{nz} -\gamma^{1/2} Y ) P = \partial_Y \gamma 
Y P+ \partial_z^2 P+\gamma \partial_Y^2 P+2 \gamma^{1/2} \partial_z \partial_Y P\,.
\]

At lowest order in $\gamma$ we get
\beq
\partial_z (\mu -  e^{nz} ) P_0 - \partial_z^2 P_0 =0\,,
\eeq
that leads to
\[
P_0 = A(Y) \exp[\mu z - e^{nz}/n ] \equiv A(Y) \Pi(z)\,.
\]    
Note that $\Pi$ is the p.d.f. obtained when the noise is white, i.e. associated to  on-off behavior.
At next order we have
\beq
\partial_z (\mu -  e^{nz} ) P_1 - \partial_z^2 P_1 = \partial_z ( Y P_0 +2 \partial_Y P_0 ).
\label{sxs} 
\eeq
The solvability condition (here integration over $z$) does not set the  amplitude $A(Y)$  at this order, however eq. (\ref{sxs}) can
be easily solved to obtain $P_1$.

At second order we have
\beq
\partial_z (\mu -  e^{nz} ) P_2 - \partial_z^2 P_2 = \partial_z ( Y P_1 +2 \partial_Y P_1 ) + \partial_Y Y P_0 + \partial_Y^2 P_0. 
\eeq
Integrating over $z$, we obtain
\beq
0=\partial_Y ( A Y  +  \partial_Y A  ) \,,
\eeq
that leads to 
\beq
P_0 = \exp[\mu z - e^{nz}/n - \gamma y^2/2  ] 
\eeq
Using this result to calculate the moments for small $\mu$, we recover  the on-off scaling
$\beta_m=1$ for all moments $m$. The validity of this expansion however holds
only when $\gamma$ tends to zero with fixed  $\mu$ and thus it does not provide the actual critical exponent.
 The two limits of small $\mu$ and small $\gamma$ cannot be exchanged:
\[
\lim_{\gamma\to0} \beta_m 
\ne \lim_{\mu \to 0} \lim_{\gamma \to 0}\log (\langle x^m \rangle) / \log(\mu).
\]

This explains the crossovers that are observed if we calculate the moments numerically for small values of $\gamma$.  
In fig. \ref{fig3}, we display the first moment as  a function of $\mu$ for $n=2$. For large $\gamma$, the mean-field exponent $\beta=1/2$ 
is found for $\mu$ up to unity. In contrast for small $\gamma$, this exponent is recovered only for very  small $\mu$. For larger $\mu$, 
an apparent  exponent $\beta$ smaller than unity  can be estimated which traces back to the on-off behavior and is only a cross-over. 
We note that in an experiment or in a numerical simulations, such cross-overs can easily be interpreted as anomalous exponents. 
Indeed  they can be observed on a large range of $\mu$ and  the deterministic result is only recovered for very small $\mu$.

\section{Anomalous exponents for a less steep potential}
\label{Sec_Anm}

\begin{figure}[h!]
\begin{center}
\includegraphics[width=8cm]{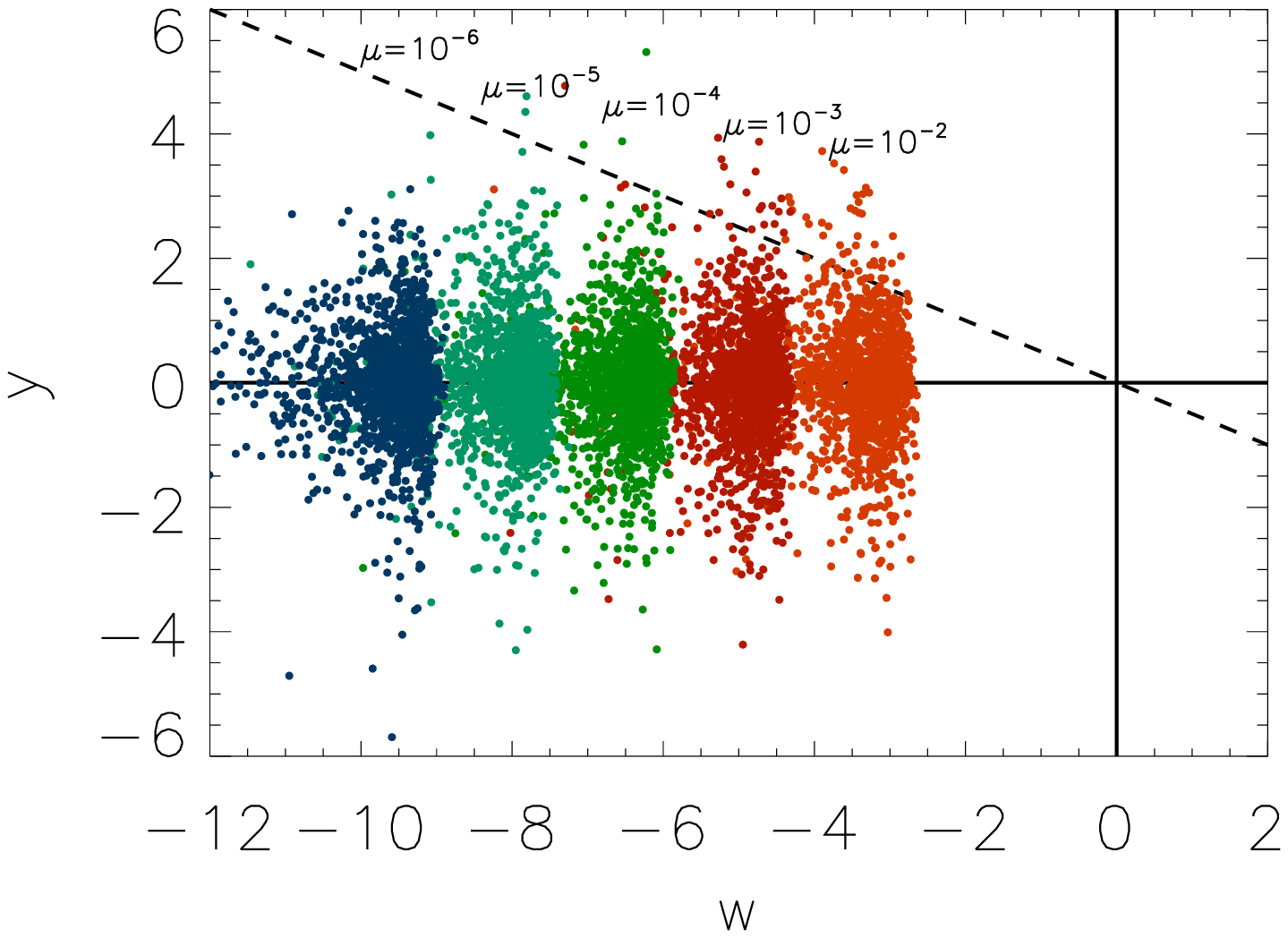}
\includegraphics[width=8cm]{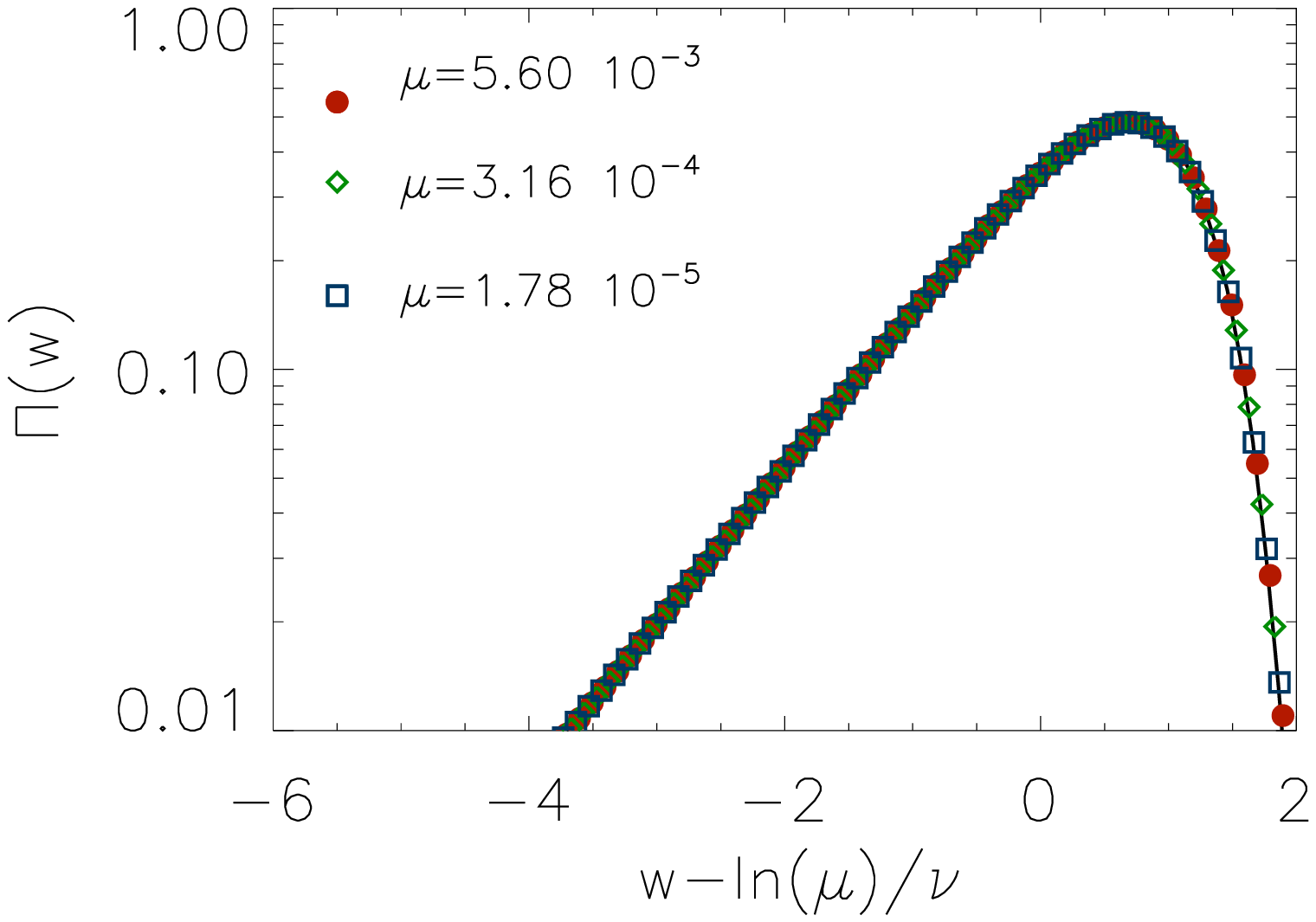}
\caption{\label{fig4}  
                      Top panel: Position in phase-space of  $10^3$ realizations of the numerical solution 
                      of the Langevin Equation \ref{eqdepart} with $F=\nu |y|$, and $\nu=1$.  
                      Different shades/colors (online) correspond to different values of $\mu$ as indicated. The dashed line corresponds to $w+y=0$ where the nonlinear term is of order one.  
                      Lower panel: The marginal probability $\Pi(w)=\int P(y,w)dy$ from the numerical investigation
                      (dots) and the analytical prediction (solid line) 
}
\end{center}
\end{figure}

The  expansion presented in section IV.A fails when the denominator in eq.  (\ref{eqflux}) diverges.
A simple potential $F$ for which this expansion can break down is $F = \nu |y|$. 
In this case $y$ follows a Brownian motion with solid friction \cite{touchette}. 
Time series of $x$, displayed in fig. \ref{figtime}, show an intermediate behavior between on-off intermittency and the one described in  section III.B. Phase-spaces for the numerically computed solutions of eq. (\ref{eqdepart}) are displayed in fig. \ref{fig4}.  

For this choice of $F$, we need to consider two cases separately. For   $\nu < n$, the expansion of section III breaks down at  lowest order and a different asymptotic must be performed. For $\nu > n$, the expansion remains valid for the first moments but need to be modified for larger moments. 

\subsection{ The case $\nu > n$ } 
In this case the previous expansion works only for the calculation of the moments
of order smaller than $\nu$. To fix this we need to consider a separate expansion for small and for large $y$.
For $y=\mathcal{O}(1)$ the previous expansion is valid and the probability distribution function
in this region is given by 
\beq
P^{in} \simeq \sqrt{ \frac{  \delta \nu^2  }{4 \pi \mu }}  \quad    e^{-\delta u^2 -\nu |y|}\,,
\label{solin}
\eeq
with $\delta$ given by equation (\ref{eqlambda}).
This solution is sufficient to calculate the moment $m$
provided that  $m<\nu$. Indeed, we have
\beq
\langle x^m \rangle  = \int e^{m(y+w)} P dy dw = \frac{\nu}{\nu^2-m^2} X_0^m \mu^{m/n}\,.
\label{eq34}
\eeq 
Thus in this case, we obtain the deterministic scaling 

\beq 
\beta_m = m/n  \quad \hbox{for} \quad m<\nu \quad \hbox{and} \quad n<\nu.
\label{betaAa}
\eeq

For $m>\nu$ the integral in equation (\ref{eq34}) diverges and the calculation for the moments fails because the expansion does not capture 
the large $y$ behavior of the pdf. To remedy this we need to calculate the large $y$ behavior of the pdf 
$P^{out}$.
We then rescale variables to $y=y'-\ln(\mu)/n$ and $w=w'+\ln(\mu)/n$ (so that $y'+w'={\mathcal{O}}(1)>0$), and obtain 
\beq
\partial_{w'} (e^{n(w'+y')}  P^{out}) + \nu  \partial_{y'}  P^{out}  + \partial_{y'}^2 P^{out} =0\,.
\eeq
Making the  change of variables  $P^{out}=e^{-\nu y' -n w'} \Theta$ and $\tau=e^{-nw'}$, $r=e^{ny'}/n$, we arrive at
\beq
          \partial_\tau  \Theta  =  - \tilde{\nu}        \partial_r  \Theta +       \partial_r r \partial_r  \Theta
\eeq
which is an advection-diffusion equation with space-varying diffusivity.
The advecting velocity $\tilde{\nu}=\nu/n$ is directed away from the $r=0$ boundary.
The boundary conditions  are $\Theta \to 0$ for $r^2+\tau^2 \to \infty$ and $\Theta = f(\tau)$
for $r\to0$ and $\tau$ finite, where $f(\tau)$ is determined by matching with the inner solution.
This problem can be solved exactly and its solution is given by
\[
\Theta (\tau,r) = \frac{1}{\Gamma(\tilde{\nu})}  \int_{-\infty}^\tau  f(\tau_0) \frac{r^{\tilde{\nu}}}{(\tau-\tau_0)^{{\tilde{\nu}}+1}} e^{-r/(\tau-\tau_0)}  d\tau _0.
\]
To obtain the functional form of $f$ we
match at an intermediate value of  $y$ chosen to be $\alpha/n\ln(\mu)$ with $0<\alpha <1$. We obtain that to first order in $\mu$
\[ f(e^{-w'}) \simeq  \mu^{\tilde{\nu}            } \,\quad        \frac{          \nu    }{2      }   \delta(w'-w_0) \,, \]
thus
\beq
P^{out}(y',w') = \frac{\nu n \,\, \mu^{\tilde{\nu}}}{2 n^{\tilde{\nu}}  \Gamma(\tilde{\nu})}  \frac{e^{-nw'} \exp\left[ \frac{ -e^{ny'}}{n(e^{-nw'} - e^{-nw_0})} \right]   }{(e^{-nw'} - e^{-nw_0})^{{\tilde{\nu}}+1}} 
\label{solout}
\eeq
for $w'<w_0$ and zero otherwise.
The calculation of the higher moments $m>\nu$ can then be performed as
\[  
    \langle x^m\rangle  = \left[ \int^{+\infty}_{-\infty}  \int^{-\frac{\alpha}{n}\ln(\mu)}_{-\infty}  e^{my+mw} P^{in}(w,y) \,dy \,dw  \right] + \]
\[        \qquad
                          \left[ \int^{+\infty}_{-\infty}  \int^{+\infty}_{-\frac{\alpha}{n}\ln(\mu)}  e^{my+mw} P^{out}(w,y) \,dy \,dw  \right] \,. \]  
In the limit $\mu\to0$ and for $m>\nu$ the main contribution comes from the second integral resulting in
\beq
\langle x^m\rangle \quad = \quad \frac{\nu \Gamma(\tilde{m})X_0^{\nu} J(\tilde{m},\tilde{n})}{2 \Gamma(\tilde{\nu}) n^{\tilde{\nu}-\tilde{m}+1 }} \quad \mu^{\tilde{\nu}}
\eeq
where  $J(m,\nu)=\int_0^\infty (u+1)^{-m}u^{m-\nu-1}du$. 
Thus the scaling exponent $\beta_m$  of the m-th moment is
 
\beq \beta_m = \nu/n  \quad \hbox{for} \quad m>\nu  \quad \hbox{and} \quad \nu>n. 
\label{betaAb}
\eeq

\subsection{ The case $\nu < n$ }

When $\nu < n$ the previous expansion fails due to the divergence of the denominator in eq. (\ref{eqflux}).
 Figure \ref{fig4} shows   the location in phase space of $10^3$ trajectories for different values of $\mu$ (different colors) obtained
by the numerical integration of the Langevin equations.
It can be seen that  
as $\mu$ becomes smaller the distribution moves to smaller values of $w$ but retains its width, unlike the case of the steep potential.
We thus need a different expansion. Making the substitution 
$w=w'+\frac{1}{n}\ln(\mu)$,  the Fokker-Planck  equation becomes
\beq
\mathcal{L}_0 P =
\mu \partial_w' \big{[}(1 -  e^{n(w'+y)})P \big{]}  
\eeq
Since the derivative with respect to $w'$ is multiplied by the small parameter, we write
\[
P = \exp\left( \frac{1}{\mu} S(w') \right)    R(w',y)\,,
\] 
and expand $R(w',y)$ as  $R=R_0(w',y)+\mu R_1(w',y)+\dots$. 
At lowest order we obtain 
\beq
{\cal L}_\sigma R_0 \equiv \sigma ( e^{n(w'+y)}-1 ) R_0 +  \mathcal{L}_0 R_0 = 0.
\label{0thor}
\eeq
where $\sigma = dS(w')/dw'$.
This equation can be solved exactly for positive and negative $y$.
The two solutions are then matched at $y=0$ that  selects the value of $\sigma$
\begin{equation}
n=2\nu I_\kappa \left[\lambda e^{nw'/2}\right] K_\kappa\left[ \lambda e^{nw'/2}\right]\,,
\label{lambdaeq}
\end{equation}
where $\lambda^2 = - 4\sigma /n^2$, $\kappa=\sqrt{\nu^2/n^2-\lambda^2}$
and $I_\kappa$ and $K_\kappa$ are modified Bessel functions of order $\kappa$.

We can find an approximate solution of Eq. (\ref{lambdaeq}) to obtain $\sigma$
for $w'\rightarrow -\infty$.
To proceed we use the following relation for Bessel functions 
$I_a\left[z\right]K_a\left[z\right]\simeq 1/(2a)+ C_1 z^{2 a}+C_2 z^2+\dots$, valid for $z\rightarrow 0$ 
and where $C_1$ and $C_2$ are two constants. In this limit and for $\nu < n$, we obtain 
$\lambda \propto e^{n w' \nu/(2(n-\nu))}$.
The asymptotic behavior of $S$  is then of the form 
\beq 
S \simeq - \frac{n (n-\nu)}{4 \nu} e^{n w' \nu/(n-\nu)}. \label{negw} 
\eeq 
We  observe that the exponential term $e^{S(w')/\mu}$ acts as a cut-off that selects very negative values 
of $w'$ in the small-$\mu$ limit. This limit  will turn out to be useful when we calculate the moments.
For the time being we proceed with our expansion without considering the $w'\rightarrow -\infty$ limit.

The solution for $R_0$ is  $R_0=A(w') e^{-F/2} g(w',y)$ where $g$ is defined by 
\begin{equation}
g \equiv 
\left\{
\begin{array}{l}
  K_\kappa\left[\lambda e^{n(w' +y)/2}\right]\,I_\kappa\left[ \lambda e^{n w'/2} \right]        \quad ({y > 0}) \\ \, \\
  I_\kappa\left[\lambda e^{n(w' +y)/2}\right]\,K_\kappa\left[ \lambda e^{n w'/2} \right]        \quad ({y < 0}) \, .
\end{array}
\right.
\end{equation}
The solution $R_0$ decays exponentially $R_0\sim e^{\nu y}$ for $y\to-\infty$ while for positive $y$ the exponential decay
($R_0\sim e^{-\nu y}$) is followed by a super-exponential cut-off ($R_0\sim \exp[-e^{ n y}]$) for $y \gg -w'/(1-\nu/n)$. 
The amplitude $A(w')$ is obtained by a solvability condition on the equation at  next order 
\beq
{\cal L}_\sigma R_1=\partial_{w'} [(1-e^{n(y+w')})R_0]\,.
\label{eqord1}
\eeq 
To obtain the solvability condition we need to multiply and integrate Eq. \ref{eqord1} by $e^{F} g(w',y)$ that is
an element of the kernel of the adjoint operator of ${\cal L}_\sigma$ and thus the left hand side integrates to zero. 
The resulting solvability condition after some transformations reads 
\begin{equation}
\partial_{w'} \log{\left(A \sqrt{ | e^{n w'} \mathcal{I}_n -\mathcal{I}_0|}\right)} =-\frac{n e^{n w'} \mathcal{I}_n}{2( e^{n w'} \mathcal{I}_n -\mathcal{I}_0)}\,,
\label{eqA}
\end{equation}
where $\mathcal{I}_q(w')$ is the integral 
\[
\mathcal{I}_q = \int_{-\infty}^{\infty} e^{qy} g^2(w',y) dy\,.
\label{defIq}
\]
%
%
We can find the asymptotic behavior of $\mathcal{I}_0$  and $\mathcal{I}_n$ in the limit $w'\to \infty$ that leads to
$\mathcal{I}_0 \simeq n^2/2\nu^3$ and
$\mathcal{I}_n \simeq (n/2\nu^2)e^{-nw'}$. Inserting this result into Eq. \ref{eqA},  we obtain that 
\[A(w') \simeq \exp\left[\frac{n\nu}{2(n-\nu)}w'\right]  \]
for $w'\rightarrow -\infty$. 
Already at this point we can observe by balancing the behavior of $A$  with the expression for 
$S$ in eq. \ref{negw}, that the most probable $w$ scales like $w \propto \frac{1}{\nu} \ln(\mu)$ that is different
 from the scaling $w \propto \frac{1}{n} \ln(\mu)$ observed in section IV A. A comparison of the  asymptotic result for the pdf with 
the results of numerical simulations can be seen in the lower panel of figure \ref{fig4}.


We can now calculate the moments
\begin{equation}
\langle x^m\rangle=\langle e^{m w +m y} \rangle=\mu^{m/n} \langle e^{m w'+m y} \rangle=\mu^{m/n} M_m/M_0
\end{equation}
 where  we have introduced 
\begin{equation}
M_q=\int_y \int_{w'} R_0(w',y) e^{q w'+q y} e^{S(w')/\mu} dw' dy\,.
\end{equation} 
 
The small $\mu$ behavior of $M_q$ is obtained by keeping in mind that we can restrict to 
very negative values for $w'$ which simplifies the expression of the Bessel functions. 
For $m>\nu$, the major contribution of the integral $M_m$ comes from the large $y$ 
and large $w$ 
part of the pdf that scale like $y \sim -w  \sim -1/\nu  \ln(\mu)$.
For $m<\nu$ the major contribution comes from the small $y\sim 0$  and large $-w\sim \frac{1}{\nu} ln(\mu)$. 
Careful evaluation of the integrals in the  $\mu\to0$ limit then leads to  
\begin{equation}
\beta_m=\hbox{min} \left[\frac{m}{\nu},1\right]\, \quad \hbox{for} \quad \nu < n.
\label{betaB}
\end{equation}

\subsection{General expression for $\beta_m$} 

\begin{figure}[h!]
\begin{center}
\includegraphics[width=8cm]{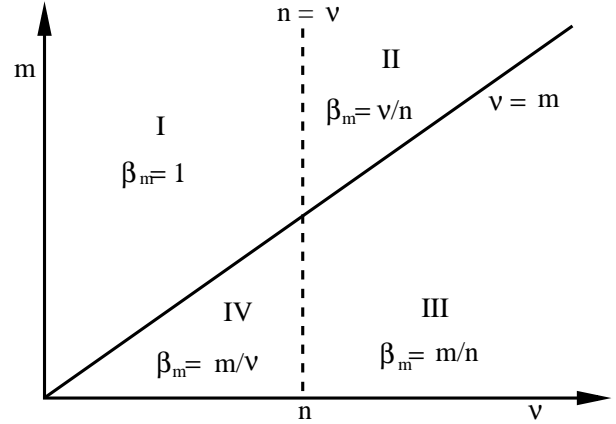}
\caption{\label{fig5}  Phase diagram displaying the four regions where different scalings  of the moments $\beta_m$ are observed.  }
\end{center}
\end{figure}

\begin{figure}[h!]
\begin{center}
\includegraphics[width=8cm]{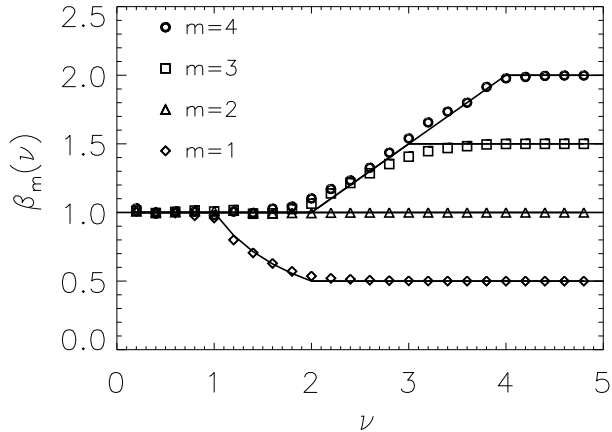}
\caption{\label{fig6} Critical exponents of the first 4 moments ($\beta_1$ diamonds, $\beta_2$ triangles, $\beta_3$ squares, $\beta_4$ circles.) 
as a function of $\nu$ for $n=2$. The results were obtained from numerical integration and are compared with the analytical predictions. }
\end{center}
\end{figure}

We have obtained several analytical expressions for the critical exponents that depend on the values of $m$, $n$ and $\nu$. 
These expressions given in Eqs. (\ref{betaAa},\ref{betaAb},\ref{betaB}) can be written in a compact form 
\begin{equation}
\beta_m=\frac{\hbox{min}[m, \nu]}{\hbox{min}[n, \nu]}\,,
\label{betatot}
\end{equation}
that divides the $(\nu, m)$-parameter space in four distinct regions with different scaling behaviors, 
as displayed  in fig. \ref{fig5} and fig. \ref{fig6}.
Depending on the statistical measure examined (degree of the moment $m$) different transitions can be observed.
Increasing $\nu$ we observe that 
the system transitions from on-off behavior  $\beta_m=1$ (when $\nu$ is small and the noise is dominated 
by the $\delta$-correlated component) to the anomalous scaling $\beta_m=m/\nu$ (for $m<n$) or $\beta_m=\nu/n$ (for $m>n$) 
and finally to the deterministic one (when $\nu$ is large and the distribution of $y$ is narrow). 
We refer to the two intermediate scalings (for $\nu$ between $m$ and $n$) as anomalous because they do not follow neither the mean field 
nor the on-off prediction. We note that the exponents are continuous functions of the parameter $\nu$ but 
not analytic. Therefore, they cannot be captured as a single Taylor series valid over the whole parameter space. 

For moments of small degree, the expression for the anomalous exponent $m/\nu$ does not involve the nonlinearity $n$ while for moments of 
large degree, the exponent $\nu/n$ does not depend on the considered moment. 

For fixed value of  $\nu$ and of the nonlinearity $n$, the scaling of the moments with $m$ contains two regimes. A linear behavior 
for small $m$ and a plateau for large $m$. The value at the plateau depends on the width of the noise: corresponding to on-off ($\beta=1$) 
for wide noise (small $\nu$), and a different value $\nu/n$ for a narrow noise (large $\nu$). 
We point out that the moments do not depend linearly on $m$, {\it i.e.} that the solutions display multiscaling. This traces back to the non-trivial expressions found for the p.d.f. obtained in section V.

In figure \ref{fig6} we display the first 4 exponents measured from numerical simulations. 
To obtain these exponents the Langevin equations were solved for $\mu$ in the range $10^{-6}<\mu<10^{-3}$ and for 
a duration long enough for the 4th moment to be converged. 
The exponents were then calculated by a linear fit. 
Simulations with $\mu$ as small as $10^{-7}$ for which only the first two moments were converged 
were also performed to verify that transient behavior as the one observed in section IV.C is not present.
The measured exponents are in agreement with the results obtained analytically.

\section{A heuristic determination of the critical exponents}
\label{Sec_Hrs}

\begin{figure}[h!]
\begin{center}
\includegraphics[width=8cm]{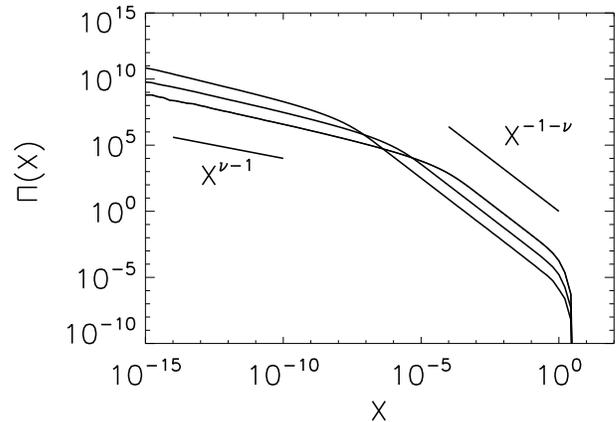}
\caption{\label{figpdf}  
The marginal probability density function of $x$ , $\Pi(x)$, computed from the numerical solution of eq. \ref{eqdepart} 
for $\nu=0.4$ and for three different $\mu$ ($10^{-3}$, $10^{-4}$,  $10^{-5}$). The position of the crossover 
between the power-laws moves to smaller values when $\mu$ decreases. }
\end{center}
\end{figure}

In this section, we try to explain the predicted critical exponents by giving a physical 
interpretation of the anomalous behavior. 
The evolution of $z=\log(x)$ satisfies $\dot{z}=\mu -x^n +\dot{y}.$ 
 We note that close to criticality
($\mu\ll 1$) the amplitude of $x$ remains small most of the time and thus  the noise $\dot{y}$ is the dominant effect. 
Keeping only this effect, we have 
$\dot{z}=\dot{y}$ that leads to the relation $z= y+C$, 
where the value of the integration constant $C$ needs to 
be determined.
Accordingly the marginal probability writes $\Pi_z(z)=\Pi_y(z+C)=\frac{1}{N}e^{-\nu|z-C|}$. 
This relation is only violated at large $x=\mathcal{O}(1)$ where the nonlinearities 
need to be taken into account and  provide a large $z$ cut-off. Taking all these into
account and returning to the $x$ variable we can write  the marginal probability $\Pi_x(x)$ as 
\begin{equation}
\label{2pl}
\Pi_x(x)= \frac{1}{N}
\left\{
\begin{array}{l}
 x^{-1+\nu}x_c^{-\nu} \quad  \hbox{if} \,\, 0\,\,<x < x_c \\
 x^{-1-\nu}x_c^{+\nu} \quad  \hbox{if}  \,\, x_c<x < x_{_{NL}}   \\
\qquad 0           \qquad  \quad  \hbox{if} \,\,   x_{_{NL}}  <x    
\end{array}
\right.
\end{equation}
where $x_c=e^C$ is still undetermined, $x_{_{NL}}=\mathcal{O}(1)$ stands for  the nonlinear cut off 
and $N$ is determined by normalization. 
Figure \ref{figpdf} shows the marginal probability distribution
$\Pi_x$ obtained from the numerical integration of the Langevin equations demonstrating that
the two power laws model (\ref{2pl}) gives a qualitative good description of $\Pi_x$. 
Straightforward estimates of the moments $\langle x^{n}\rangle=\int x^{n} \Pi_x dx $ for this form of pdf result in 
\begin{eqnarray}
\langle x^m\rangle & \propto           &  \qquad \,          x_c^m    \,\,\, \hbox{if\,\, $  m \le \nu$}     \nonumber\\
                   & \propto           & x_{_{NL}}^{m-\nu}   x_c^\nu  \,\,\, \hbox{if\,\, $\nu \le   m$} \,. \label{eqheuristic}
\end{eqnarray}
 
 All that is left is to determine the value of $x_c$. 
 This can be obtained by  balancing the averaged effect of the nonlinearity  ($\langle x^n \rangle$) with the linear drift ($\mu$).   
As the estimates for the moments indicate two cases need to be considered. 

For $\nu>n$, the average value of the nonlinearity is given by $\langle x^{n}\rangle \propto x_c^{n}$. 
Balancing with the drift leads to $x_c\sim \mu^{1/n}$. 

 For $\nu<n$, the tails of the pdf and the large $x$ cut-off determine 
the averaged value of the nonlinearity  given by $\langle x^{n}\rangle \propto x_c^{\nu}x_{_{NL}}^{n-\nu}\sim x_c^{\nu}$. 
Balancing with the drift leads to $x_c\propto \mu^{1/\nu}$. 

Inserting the expressions of $x_c$ into eq. (\ref{eqheuristic}), we obtain all the behaviors predicted by eq. (\ref{betatot}).
 It is clear from these arguments that the tails of the $y$ distribution control the presence of anomalous exponents. 
If $y$ has a narrow distribution ($F$ is a steep potential) $x$ does not deviate far from the deterministic value
and thus mean-field scaling is obtained.
On the other hand, if  $y$ has a wide distribution it is the tails of the pdf, and the rare visits of $x$ to the nonlinear 
regime $x\sim x_{_{NL}}$ that determine the balance with the nonlinearity 
and the resulting scaling is anomalous.

\section{Conclusions}

We have introduced and studied the behavior of  a family of zero-dimensional models in the vicinity of the instability threshold. 
The amplitude of the unstable mode in our models evolve in the presence of multiplicative fluctuations that makes it possible to have bifurcation at zero dimensions. 
Because this model is zero-dimensional (does not depend on space), it is among the simplest that can be considered.
However, despite its simplicity the model exhibits nontrivial behavior. 
Depending on the control parameters 
the solutions display anomalous scaling close to the onset of instability: 
moments scale with the distance to onset as power-laws different than the ones predicted by mean-field theory.
In addition, the system displays multi-scaling: the exponents  are not simply proportional to the moment degree. 
 
The values of the exponents were obtained by an exact calculation through perturbative expansions in the departure from criticality. 
This differs from what is usually obtained in equilibrium phase transitions where the exponents are expressed as a series in the 
spatial dimension minus the critical dimension. In addition to the expansion we have presented  heuristic arguments that allow
to determine the critical exponents. This enables us to identify the basic ingredients for obtaining the anomalous behavior.
First the model relies on a noise which spectrum vanishes at zero frequency. 
We note that in the context of advection of a passive scalar in a turbulent flow, stochastic processes that have
vanishing spectrum at zero frequency have been used to model  anomalous diffusion \cite{bill}.
In the present model vanishing spectrum at zero frequency is achieved by considering the derivative 
of a noise $y$ that follows a random walk in a confining potential. If $y$ has a narrow distribution, normal scaling is obtained. 
If $y$ has a  wide distribution, truly anomalous behavior takes place.

All the analytical results were tested and verified by numerical simulations. Convergence of the estimated exponents from 
the numerical results proved significantly difficult due to the appearance of intermediate power laws that are present
in some limiting cases. These intermediate power laws can contaminate the value of the exponents if sufficiently small values of
$\mu$ are not investigated. Experimentally if such cross overs exist it may be difficult to distinguish them from 
true anomalous scaling, given the experimental limitations.

As an example we mention that  in a recent experiment, the dynamo instability was observed in a turbulent flow of liquid sodium \cite{VKS}, \cite{GAFD}.  
The first moment displays an exponent $0.78$ located in-between $1/2$  (as expected for cubic nonlinearities) and $1$ (as expected for on-off intermittency).   
The present model  gives a possible explanation for the observed exponent since $\beta_1$ can be larger than the mean-field $1/n$ prediction 
and smaller than the on-off exponent $1$.
However the dynamo equations are more complicated than the model considered here: two fields 
(magnetic field and velocity field) are coupled and depend on space and time. Whether and when the dynamo problem can be reduced to the model 
studied here remains an open question. 

Several additional investigations can be thought of. Other critical exponents can be defined and studied. The response to a constant field 
and the associated susceptibility are of particular interest. It is expected that simple scaling relations between the exponents exist (as in 
equilibrium phase transitions) and can be obtained exactly in the present context. 
Finally, taking space into account is also an attracting path. A possible attempt being to search for an expansion in the dimension since we have 
obtained a solution in the $d=0$ case.

\acknowledgments

The authors would like to thank Stephan Fauve for his support and suggestions.
Numerical computations were performed using the MESOPSL parallel cluster, 
the new supercomputing center of Paris Science and Letters, and their computational support is acknowledged.
 



\begin{thebibliography}{5}
\bibitem{wien} By virtue of the Wiener-Kintchin theorem, the noise spectrum is the Fourier transform of the autocorrelation function. We thus recognize here the noise spectrum at zero frequency.
\end{thebibliography}
\end{document}